\documentclass[fleqn,10pt]{wlscirep}
\usepackage{etex}

\usepackage{amsmath,amsfonts,amssymb, tikz, tikzpfeile, verbatim, wrapfig, graphicx, array, subfig, caption, float, enumitem, setspace, tabulary, epstopdf}

\usetikzlibrary{arrows,positioning,shapes} 

\newcolumntype{C}[1]{>{\centering\let\newline\\\arraybackslash\hspace{0pt}}m{#1}}

\newcounter{term}
\renewcommand*{\theterm}{\Alph{term}}

\AtBeginDocument{%
  \let\mylabel\label
}

\newcommand{\mytag}[1]{%
  \begingroup 
    \refstepcounter{term}%
    \mylabel{#1}%
    \text{\theterm}%
  \endgroup
}

\title{Theoretical Kidney Model Preprint}
\title{Modeling the Effects of Multiple Myeloma on Kidney Function}

\author[1,*]{Julia C. Walk}
\author[2,3,4]{Bruce P. Ayati}
\author[5]{Sarah A. Holstein}
\affil[1]{Concordia College, Department of Mathematics, Moorhead, 56560, USA}
\affil[2]{University of Iowa, Department of Mathematics, Iowa City, 52242, USA}
\affil[3]{University of Iowa, Program in Applied Mathematical and Computational Sciences, Iowa City, 52242, USA}
\affil[4]{University of Iowa, Department of Orthopaedics and Rehabilitation, Iowa City, 52242, USA}
\affil[5]{University of Nebraska Medical Center, Division of Oncology \& Hematology, Omaha, 68105, USA}

\affil[*]{jwalk@cord.edu}


\begin{abstract}
Multiple myeloma (MM), a plasma cell cancer, is associated with many health challenges, including damage to the kidney by tubulointerstitial fibrosis. We develop a mathematical model which captures the qualitative behavior of the cell and protein populations involved. Specifically, we model the interaction between cells in the proximal tubule of the kidney, free light chains, renal fibroblasts, and myeloma cells. We analyze the model for steady-state solutions to find a mathematically and biologically relevant stable steady-state solution. This foundational model provides a representation of dynamics between key populations in tubulointerstitial fibrosis that demonstrates how these populations interact to affect patient prognosis in patients with MM and renal impairment.

\end{abstract}

\begin{document}

\flushbottom
\maketitle

\thispagestyle{empty}

\section*{Introduction}

Multiple myeloma (MM) is a plasma cell cancer causing development of bone disease characterized by severe bone pain and bone fractures. Other associated health challenges include hypercalcemia, anemia, and kidney damage. The American Cancer Society predictions for the United States in 2018 include 30,770 new cases of MM and attribute about 12,770 deaths to MM. \cite{ACS2018} Most cases of multiple myeloma are diagnosed in older populations; the median age at diagnosis is 70, and only 5-10\% of patients with MM are under 40 years old, with less than 1\% under 35 years old. \cite{Durie2007, Gertz2014, ACS2018} 

Thus far, mathematical modeling related to multiple myeloma has focused on the breakdown in bone remodeling process caused by malignant plasma cells. \cite{Ayati2010, Wang2011, Ji2014} In this paper, we focus instead on the kidney damage that occurs in some patients with MM because of the significant effects of kidney dysfunction on patient prognosis. 

Several studies have reported inferior overall median survival time for patients who present with renal impairment, which occurs in approximately $50\%$ of patients with MM.\cite{Yadav2015} A study by Knudsen et al. in 2000 grouped patients according to degree of renal impairment: none (defined as creatinine $< 1.47$ mg/dL), moderate (creatinine 1.47-2.26 mg/dL), and severe (creatinine $> 2.26$ mg/dL). They found a median survival of 36 months for patients in the first group, 18 months for patients in the second group, and 13 months for patients in the third group. Furthermore, the study concluded that reversibility of renal failure slightly improved prognosis.\cite{Knudsen2000} More recently, a 2014 retrospective study that included patients from 15 Swedish hospitals used estimated glomerular filtration rate (eGFR) to study effects of renal impairment on overall survival time with two types of treatment. Overall median survival time (33 months) for patients presenting with renal impairment (defined as eGFR $<$ 60 mL/min/$1.73 \text{ m}^2$) was significantly less than overall median survival time (52 months) for patients with no renal impairment.\cite{Uttervall2014} A 2015 study by Gonsalves et al. that found the median survival rate for patients with renal insufficiency (creatinine $> 2$ mg/dL) was 42 months while median survival time for patients with normal renal functions was over twice that, at 99 months.\cite{Gonsalves2015} Patients in this study were treated with one or more novel agents, and it was found that patients who demonstrated improved renal function after treatment had improved prognoses compared to patients with no response. However, the improved survival rates were significantly lower than the prognosis of patients with normal renal function. This disparity between expected survival time in patients with and without renal insufficiency is the motivating clinical problem for our model.

We present a continuous and deterministic model of dynamics between the proximal tubule cell population and free light chains in the kidney of a healthy patient and then a model that incorporates the effect of malignant plasma cells in patients with MM. We analyze our model for steady-state solutions to find a mathematically and biologically significant stable steady-state solution and find computational results for the model incorporating malignant plasma cells. Our model captures the main players and relationships in the process of renal interstitial fibrosis caused by MM and is meant to translate some basic understanding of MM for future work on predictive and prognostic models useful for patient-specific medicine.

\subsection*{Renal Function}
\begin{figure} [H]
\begin{center}
\includegraphics[width=0.4\textwidth]{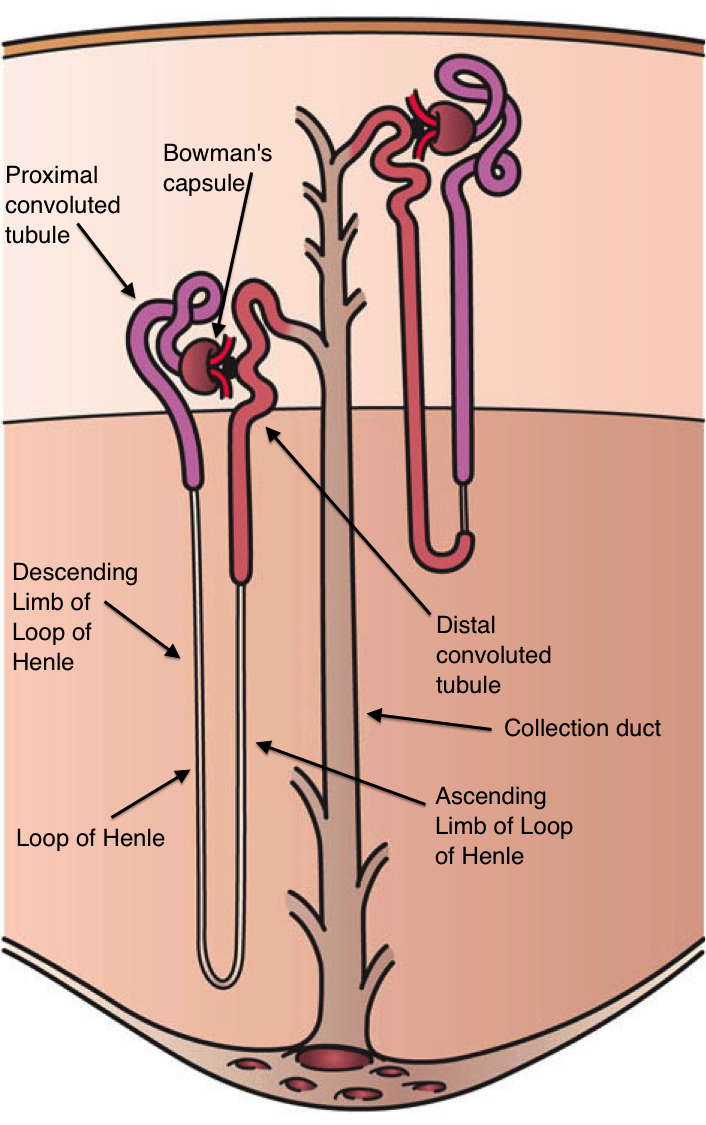}
\end{center}
\caption{Nephron anatomy. The nephrons are made up of tubules, and the proximal tubule cells in our model line the proximal convoluted tubule immediately after the blood is filtered through Bowman's capsule. We will model the amount of proximal tubule cells lining the proximal convoluted tubule throughout the process of proximal tubule cell injury via renal interstitial fibrosis in patients with MM. Image Source: https://en.wikipedia.org/wiki/Nephron.  Reproduced without modification under Creative Commons License CC BY 3.0, http://creativecommons.org/licenses/by/3.0/.  Artwork by Holly Fischer.}
\label{fig:Kidney_Nephron}
\end{figure}

In order to examine the effect of multiple myeloma on the kidney, it is necessary to first consider normal renal function. Humans have two kidneys, made up of systems of tubules, called nephrons, that are the working units of the kidney. Each kidney has approximately 1 million nephrons (Figure \ref{fig:Kidney_Nephron}). The kidneys are the body's filtering system, and help maintain the body's homeostasis. The kidneys keep substances in the human body in balance by regulation and removal of metabolic waste. 

The main functions of the kidney include regulation of water and electrolyte balance, which affects arterial blood pressure. The kidneys also monitor the excretion of hormones and the blood levels of prescription drugs that affect body function, as well as regulating red blood cell production and vitamin D production. Most of the kidneys' work involves transportation of water and solutes between the blood and filtration systems in the kidney. Any substance not transported back to the blood is excreted in urine.

Renal disease in patients with MM is usually present as proteinuria (excess serum proteins in the urine). \cite{Batuman2007,Korbet2006} In general, kidney diseases related to multiple myeloma result from the kidneys' reduced ability to properly filter substances. There are two types of common tubular kidney damage that we will consider: proximal tubular cell injury, and cast formation. The sources of this damage are malignant plasma cells, which produce monoclonal proteins called monoclonal immunoglobulin (Ig) light chains (also called M proteins). Ig light chains are subunits of antibodies, which are produced by plasma cells to fight infection, and antibodies express one of two types of light chain: kappa light chain or lambda light chain. Monoclonal myeloma plasma cells will overproduce one specific type of Ig light chains. The most common cause of severe renal failure in patients with MM is tubulointerstitial pathology resulting from high circulating concentrations of monoclonal Ig light chains.\cite{Batuman2007} 

\subsection*{Proximal Tubule Cell Injury}

Proximal tubule cell (PTC) injury is caused by free light chain interaction with PTCs. Free light chains can be toxic to PTCs by blocking transport of glucose, amino acids, or phosphates, activating tubulointerstitial fibrosis, and causing excess free light chain endocytosis.\cite{Sanders2012} Endocytosis is the process by which cells absorb proteins by engulfing them. It is important to note that not all monoclonal free light chains are toxic to the kidneys, and in fact, many patients with high amounts of serum free light chain have no renal impairment.\cite{Yadav2015} It appears that toxicity depends on the structure of a particular individual's free light chains' 3D structure or protein folding.\cite{Batuman2007,Hutchison2011} While high amounts of light chains can be a sign of multiple myeloma, a more useful laboratory parameter is the serum kappa to lambda ratio. Normal kappa to lambda ratio is 0.26-1.65, compared to 0.37-3.1 in situations of renal impairment.\cite{Hutchison2008} When the level of either kappa or lambda light chains is very high and the other is normal to very low, the ratio is considered abnormal, which suggests the presence of clonal plasma cells.

One way free light chains can be toxic to PTCs is by activating tubulointerstitial fibrosis. Tubulointerstitial fibrosis is the process initiated by the interaction between proximal tubule cells and free light chains, which activates inflammatory pathways in the kidney. Sustained inflammation causes the excess accumulation of extracellular matrix (ECM), which is eventually replaced by scar tissue.\cite{Hewitson2009} ECM, which is made up of proteins and collagens, provides structural support to surrounding cells and most cells cannot survive unless they are anchored to the ECM. The scar tissue that replaces ECM is part of the formation of excess fibrous tissue that characterizes fibrosis. This process is considered to be largely irreversible, and leads to the loss of function of proximal tubule cells and end stage renal disease (ESRD).

Because tubulointerstitial fibrosis begins with the interaction between proximal tubule cells and free light chains, the main goal of treatment is to reduce light chain production by killing the malignant plasma cells.\cite{Korbet2006,Hutchison2011a} Initial treatment includes chemotherapy drugs, hydration, and in some cases the use of bisphosphonates to lower calcium levels. Also used as treatment is plasmapheresis, which involves removing the blood plasma from the body, treating it (reducing plasma concentration of light chains), and returning it to the body. This is similar to dialysis, which is used to remove waste from the blood. Both plasmapheresis and dialysis use machines to perform the kidneys' usual job of filtering the blood. Renal transplantation is usually not considered because of the poor prognosis of patients with MM.\cite{Korbet2006}

\subsection*{Flowchart}

To better identify the various cell populations involved in the processes we model, we present a flowchart in Figure \ref{fig:Flowchart}.

The increased monoclonal protein production by the myeloma cells leads to an increased level of free light chain molecules circulating in the blood. These free light chains are either endocytosed, or precipitated.

In the primary situation that our model addresses, the increased free light chain production leads to increased light chain endocytosis by proximal tubule cells via cubilin/megalin complex.\cite{Korbet2006} Cubilin and megalin are two endocytic receptors that play important roles in renal tubular clearance and reabsorption of proteins. They initiate receptor-mediated endocytosis, a process by which cells internalize molecules. This involves an inward budding of plasma membrane vesicles containing the monoclonal proteins with receptor sites specific to the molecules being internalized. This increased light chain endocytosis activates NF-$\kappa$ B and MAPk in the proximal tubule cells. NF-$\kappa$ B is a protein complex involved in regulating the immune system's response to inflammation, and is responsible for cytokine production. Mitogen-activated protein kinases (MAPk) direct the cellular response to mitogens and proinflammatory cytokines.

The activation of NF-$\kappa$ B and MAPk initiates the production of several different types of cytokines and growth factors by the proximal tubule cells: IL-6, CCL2, IL-8 and TGF-$\beta$. IL-6 is secreted by T-cells and macrophages to stimulate immune response, and acts as a proinflammatory cytokine. IL-8 is produced by macrophages and epithelial cells. CCL2 recruits memory T-cells and dendritic cells to inflammation sites. TGF-$\beta$ is a protein that controls cell growth, apoptosis and proliferation. These cytokines and growth factors initiate proinflammatory and fibrotic pathways, and initiate Epithelial-Mesenchymal Transition (EMT), type 2. During EMT type 2, polarized epithelial cells (such as those that line the kidney tubules, in our case, proximal tubule cells) change to assume mesenchymal cell characteristics. This allows these cells increased migratory ability to migrate to an infection site, increased resistance to apoptosis, and increased production of ECM material. This all plays a part in renal interstitial fibrosis, the sustained inflammation in proximal tubule epithelial cells. Fibrosis causes a disruption in the normal genesis and breakdown cycle of ECM, which leads to excessive ECM accumulation.\cite{Liu2004} Eventually, scar tissue replaces ECM accumulation, and causes loss of function of PTCs. Ultimately, end-stage renal failure can develop. 

In the secondary situation in our flowchart, non-endocytosed free light chains precipitate, forming solids called tubular casts within the kidney tubules. These casts are formed by the reaction of Ig light chains with Tamm-Horsfall protein. The casts partially or totally block the kidney tubules, which increases intraluminal pressure, reduces glomerular filtration rate (GFR), blood flow, and tubular clearance of the light chains, which increases serum light chain levels, resulting in a never-ending cycle. Unless the casts are removed, the result is permanent nephron loss.

Current kidney physiology modeling focuses on modeling chemical exchange between compartments in the kidney, and on modeling GFR. GFR depends on age, sex, body size, and age, and gives a good indication of how well the kidney is functioning and filtering substances in the body. To our knowledge, there is no known prior mathematical work in modeling the above process of renal tubulointerstitial fibrosis caused by multiple myeloma.

\section*{Model Development}

To create our mathematical model, we use modified power law approximations, developed by Savageau and Voit.\cite{Savageau1976, SavageauVoit1982} Power laws are useful here because they capture the nonlinearity specific to biological systems such as this one, but are comparatively easy to work with analytically. Power laws have the following form: 
$$ \frac{dX_i}{dt} = \sum_k \alpha_k \prod\limits_{j} X_{j}^{g_{ij}}-\sum_k \beta_k \prod\limits_{j} X_{j}^{h_{ij}} $$
where the first term represents growth of the $X_i$ population affected by $X_j$ populations, and the second term represents death or removal of the $X_i$ population affected by $X_j$ populations. The parameters $\alpha_k$ are growth or proliferation rates and the parameters $\beta_k$ are death or clearance rates.

Based on the biological background from Figure \ref{fig:Flowchart}, we focus on the populations of PTCs, FLCs, and renal fibroblasts for our initial model for normal dynamics, and then include the tumor cell equation for our model that simulates dynamics in a patient with MM and renal degradation.

In our simplified model of normal dynamics, the growth of PTCs is governed by its own proliferation rate and the population of PTCs decreases only through apoptosis. The growth of FLCs increases at a natural production rate and decreases by a natural renal clearance rate. The growth of renal fibroblasts increases at a natural production rate and decreases by apoptosis.

\subsection*{Model of PTC and FLC Dynamics in the Kidney without Tumor}

Using the biological background and power laws discussed above, we construct a system of ordinary differential equations (ODEs) for the PTCs and FLCs in the kidney of a healthy patient:

\begin{align}\label{eq:1}
\frac{d}{dt} P(t) &= \underbrace{\beta_P P^{g_2} \left(1-\frac{L}{L_S}\right)_+^{g_1}}_{\mytag{termA}} \ - \underbrace{\mu_P P}_{\mytag{termB}},\\
\label{eq:2}
\frac{d}{dt} L(t) &= \underbrace{\gamma_L L \left(1-\frac{L}{\hat{L}}\right)_+}_{\mytag{termC}} - \underbrace{\mu_L L}_{\mytag{termD}},\\
\label{eq:3}
\frac{d}{dt} F(t) &= \underbrace{\gamma_F F \left(1-\frac{F}{\hat{F}}\right)_+}_{\mytag{termE}} - \underbrace{\mu_F F}_{\mytag{termF}},
\end{align}

\noindent where we define $(x)_+ = \max(x,0)$ and $P(t)$, $L(t)$, and $F(t)$ are the populations of PTCs, FLCs, and renal fibroblasts. In this system, \autoref{termA} represents the proliferation rate of PTCs, where the proliferation rate of PTCs decreases in the presence of increased FLCs. Note that in a healthy patient the level of FLCs will stay approximately constant. The parameters $\beta_P$ is the PTC proliferation rate, $L_S$ is the saturation level of FLCs, and $g_1$ and $g_2$ describe the strength of FLCs on PTC growth and the strength of PTCs on its own population growth, respectively. \autoref{termB} represents a decrease in the population of PTCs due to apoptosis. In the FLC equation \eqref{eq:2}, \autoref{termC} represents the increase in number of circulating FLCs, where we define $\hat{L} = (L_{\text{min}}\gamma_L)/(\gamma_L-\mu_L)$, $L_{\text{min}}$ is the minimum number of circulating FLCs and $\gamma_L$ is the growth rate of FLCs. This entire term is defined to be positive or zero. \autoref{termD} represents the natural clearance or removal of FLCs where $\mu_L$ is the natural clearance rate. In the renal fibroblast equation \eqref{eq:3}, \autoref{termE} represents the increase in number of fibroblasts where we define $\hat{F} = (\gamma_F F_{\text{max}})/(\gamma_F - \mu_F)$ and $F_{\text{max}}$ is the maximum percentage of renal fibroblasts. As with \autoref{termA} and \autoref{termC}, this term is defined to be non-negative. \autoref{termF} represents loss of renal fibroblast population due to apoptosis.

\subsection*{Model of PTC and FLC Dynamics in the Kidney with Tumor}

To simulate the dynamics between cells in the kidney in a patient with MM and renal degradation, we add three terms to our previous system, as well as incorporate a fourth equation representing the growth of malignant plasma cells via tumor equation $T(t)$. We chose to use the differential equation form of the Gompertz model to model tumor density. The Gompertz model assumes a time-dependent growth rate and a carrying capacity for tumor load, and results in a sigmoidal shape.\cite{Gerlee2013, Sarapata2014} It has been previously used to simulate the growth of malignant plasma cells in patients with MM.\cite{Ayati2010} The new system of ODEs is shown below:

\begin{align}
\label{eq:1T}
\frac{d}{dt} P(t) &= \underbrace{\beta_P \left(1-\frac{L}{L_S}\right)_+^{g_1} P^{g_2}}_{\mytag{termG}} \ - \ \underbrace{\mu_P P}_{\mytag{termH}} \ - \ \underbrace{\gamma_F L^{g_3} P^{g_4}}_{\mytag{termI}},\\
\label{eq:2T}
\frac{d}{dt} L(t) &= \underbrace{\gamma_L L \left(1-\frac{L}{\hat{L}}\right)_+}_{\mytag{termJ}} \ + \ \underbrace{\gamma_L T^{g_5} L^{g_6}}_{\mytag{termK}} \ - \ \underbrace{\mu_L L}_{\mytag{termL}},\\
\label{eq:3T}
\frac{d}{dt} F(t) &= \underbrace{\gamma_F F \left(1-\frac{F}{\hat{F}}\right)_+}_{\mytag{termM}} \ + \ \underbrace{\gamma_F L^{g_3} P^{g_4}}_{\mytag{termN}} \ - \ \underbrace{\mu_F F}_{\mytag{termO}},\\
\label{eq:4T}
\frac{d}{dt} T(t) &= \underbrace{\gamma_T T \log \left(\frac{T_S}{T}\right)}_{\mytag{termP}}.
\end{align}

As in the previous model, we define
\begin{align*}
\hat{L} &= \displaystyle \frac{L_{\text{min}}\gamma_L}{\gamma_L - \mu_L},\\
\hat{F} &= \displaystyle \frac{\gamma_F F_{\text{max}}}{\gamma_F - \mu_F},\\  
(x)_+ &=  \max(x,0) 
\end{align*}

In this set of equations, \autoref{termI} now represents the loss of PTC due to EMT (which is precipitated by the excess FLC population). The parameter $\gamma_F$ is the fibroblast growth constant, demonstrating that as the number of fibroblasts increases, more PTCs have undergone EMT. The exponents $g_3$ and $g_4$ represent the strength of FLCs on PTC transition and the strength of PTCs on their own transition via EMT. In \eqref{eq:2T}, the new term is \autoref{termJ}, which represents the increase in the number of FLCs produced by malignant plasma cells. The parameter $\gamma_L$ represents the growth rate of these plasma cells. The exponents $g_5$ and $g_6$ represent the strength of the tumor on FLC growth and the strength of FLC on its own growth. In \eqref{eq:3T}, \autoref{termN} is the same term as \autoref{termI}, but is a source of growth in the fibroblast population. The final equation \eqref{eq:4T} is the Gompertz tumor growth equation, where $T_S$ represents tumor carrying capacity, or maximum spread of the tumor.

The parameters for the Gompertz model ($\gamma_T$, $L_T$) are scaled to a dimensionless model, e.g. maximum tumor load is set to be 100. The parameter values were taken from published work by Ayati et. al.\cite{Ayati2010} All other parameters were obtained for the model using heuristic parameter estimation to a generic response. The model is a foundation for incorporating clinical data into a quantitative predictive system, which remains future work. The parameters used to obtain the results in this letter are listed in Table \ref{table:tumorparameters}.

\section*{Results and Discussion}

\subsection*{Equilibria and Stability of the Model}
\begin{figure}[t]
\centering
\subfloat[P(0)=95]{
\includegraphics[width=.3\textwidth]{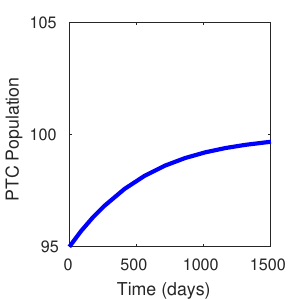}
}
\subfloat[P(0)=100]{
\includegraphics[width=.3\textwidth]{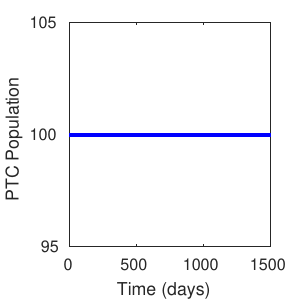}
}
\subfloat[P(0)=105]{
\includegraphics[width=.3\textwidth]{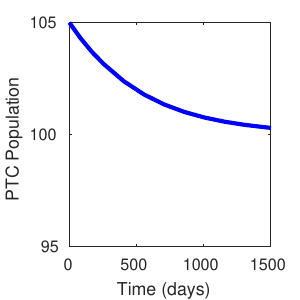}
}

\subfloat[L(0)=5]{
\includegraphics[width=.3\textwidth]{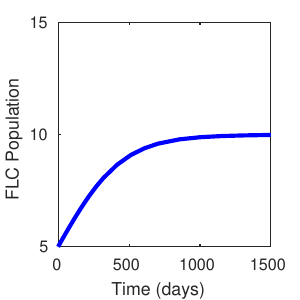}
}
\subfloat[L(0)=10]{
\includegraphics[width=.3\textwidth]{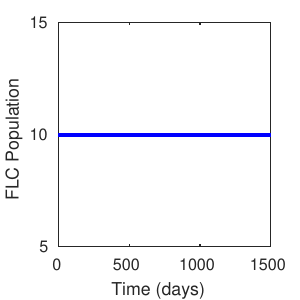}
}
\subfloat[L(0)=15]{
\includegraphics[width=.3\textwidth]{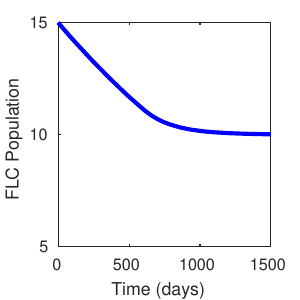}
}

\subfloat[F(0)=95]{
\includegraphics[width=.3\textwidth]{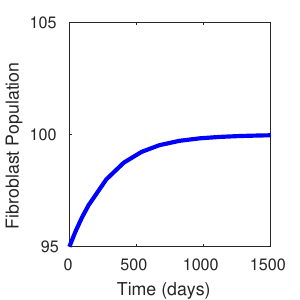}
}
\subfloat[F(0)=100]{
\includegraphics[width=.3\textwidth]{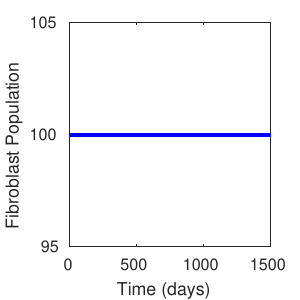}
}
\subfloat[F(0)=105]{
\includegraphics[width=.3\textwidth]{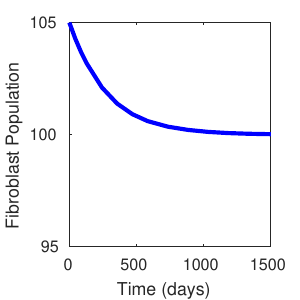}
}
\caption[Computational steady-state analysis at $(P^*, L^*, F^*) = (100, 10, 100)$]{Computational steady-state analysis at $(P^*, L^*, F^*) = (100, 10, 100)$ of PTCs, FLCs, and renal fibroblasts for Equations \eqref{eq:1}-\eqref{eq:3}, a model of behavior of pertinent populations in a patient with no myeloma cell production. These results demonstrate that in any healthy patient with $100 \%$ PTC population, $10$ mg/L FLCs, and $100 \%$ renal fibroblast population, any perturbation of these conditions (initial conditions above or below these values) will result in the body returning to homeostasis and the steady-state values, as expected. Subgraphs (a), (d), and (g) simulate initial conditions below the steady-state solution eventually returning to $(P^*, L^*, F^*) = (100, 10, 100)$. Subgraphs (b), (e), and (h), simulate initial conditions at the steady-state solution staying at $(P^*, L^*, F^*) = (100, 10, 100)$. Subgraphs (c), (f) and (i) simulate initial conditions above the steady-state solution eventually returning to $(P^*, L^*, F^*) = (100, 10, 100)$.}
\label{fig:SS_Graphs}
\end{figure}

Steady-state analysis of the model represented by Equations \eqref{eq:1}-\eqref{eq:3} yields eight steady-state solutions, summarized in Table \ref{table:genstability} with their requirements for stability based on analysis of the Jacobi matrix. We expect that healthy patients have non-zero levels of all three populations of interest, so only one of these steady-states is both mathematically and biologically significant:
\begin{align}
P^* &= \left(\frac{\mu_P}{\beta_P} \left( 1- \frac{L}{L_S} \right)^{-g_1}\right)^{\frac{1}{g_2-1}},\\
L^* &= L_{\text{min}},\\
F^* &= F_{\text{max}}.
\end{align} 
The requirements for stability of this steady-state solution are $\gamma_F > \mu_F$, $\gamma_L > \mu_L$, $g_2 < 1$. To estimate initial parameters for the model, we used known values where available and scaled the rest to maintain a stable steady-state when $P^* = 100$ for the case above; that is, a healthy patient who starts with $100 \%$ of PTC population will remain healthy. In a healthy patient, normal FLC levels are approximately 10 mg/L, so we set $L_{\text{min}} = 10$ and $L_S = 1000$, as kidney problems typically occur at FLC levels above 1000 mg/L. \cite{IMF2016} We also analyzed the stability of the steady-state $(P^*, L^*, F^*) = (100, 10, 100)$ computationally using the $\tt {ode15s}$ solver in MATLAB\textsuperscript{\textregistered}. Results of this analysis are included in Figure \ref{fig:SS_Graphs}, demonstrating that for $(P^*, L^*, F^*) = (100, 10, 100)$ our model has a stable steady-state solution and in a patient with no myeloma tumor cells present, any perturbation in normal conditions will result in eventual return to homeostasis conditions (and the steady-state solution) as expected.

\subsection*{Computational Results for Model with Tumor}
\begin{figure}[p]
\centering
\subfloat[]{
\includegraphics[width=.47\textwidth]{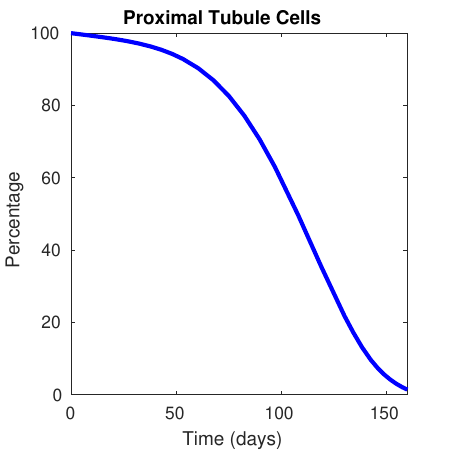}
}
\subfloat[]{
\includegraphics[width=.47\textwidth]{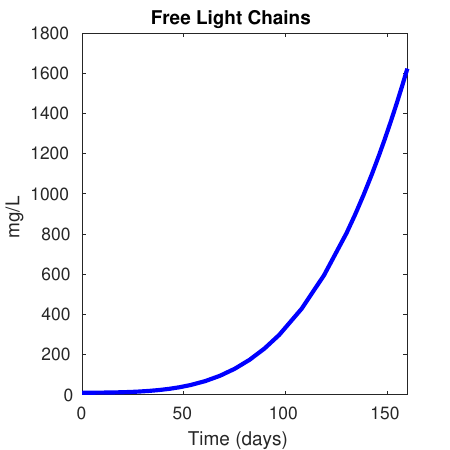}
}

\subfloat[]{
\includegraphics[width=.47\textwidth]{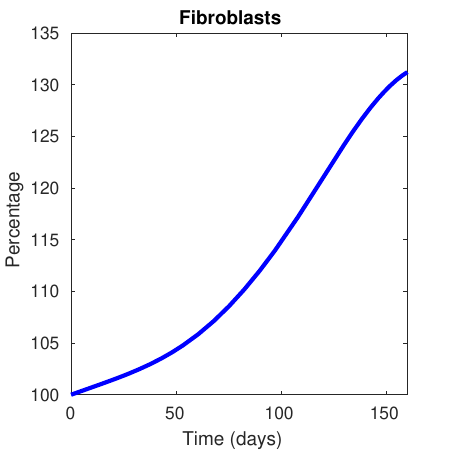}
}
\subfloat[]{
\includegraphics[width=.47\textwidth]{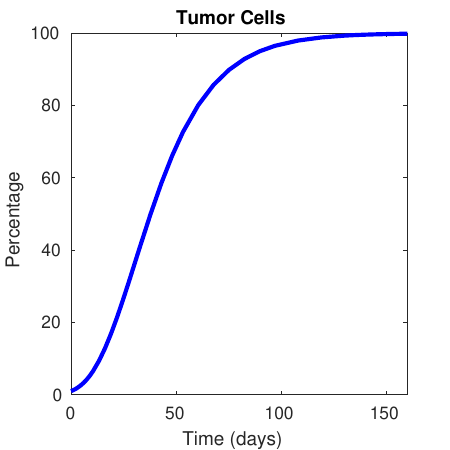}
}
\caption[Computational results for model with tumor]{Computational results for model with tumor (Equations \eqref{eq:1T}-\eqref{eq:4T}) using the parameter values in Table \ref{table:tumorparameters}, for initial conditions $P(0)=100$, $L(0)=10$, $F(0)=100$ and $T(0)=1$, over a time period of 160 days using the model given by Equations \eqref{eq:1T}-\eqref{eq:4T}. This simulation represents a fast-acting tumor that is not treated. We see that as the percentage of tumor cells increase in (d), the amount of free light chains increases beyond the threshold amount of 1000 mg/L in (b) and amount of renal fibroblasts begins to slowly increase in (c). Decline in proximal tubule cells as a response to (b), (c), and (d) is demonstrated in (a).}
\label{fig:fastTumorGraphs}
\end{figure}

Figure \ref{fig:fastTumorGraphs} shows computational results generated using  MATLAB\textsuperscript{\textregistered} {\tt ode15s}, with initial conditions $P(0)=100$, $L(0)=10$, $F(0) = 100$ and $T(0)=1$. Note that these initial conditions represent that the patient has $100\%$ of PTCs, normal levels of FLCs at 10 mg/dL and $100\%$ of fibroblast population. These three values all are consistent with a healthy patient with no myeloma cells present in the body. However, the initial tumor density begins at $1\%$ and $\gamma_T = .05$ $\text{days}^{-1}$, representing a quickly growing population of malignant plasma cells. These results are consistent with the biology described in Figure \ref{fig:Flowchart}: as the tumor cell population grows, the free light chain population increases, the proximal tubule cell population begins to decrease and the fibroblast population increases.

\begin{figure}[p]
\centering
\subfloat[]{
\includegraphics[width=1\textwidth, height=6cm, keepaspectratio]{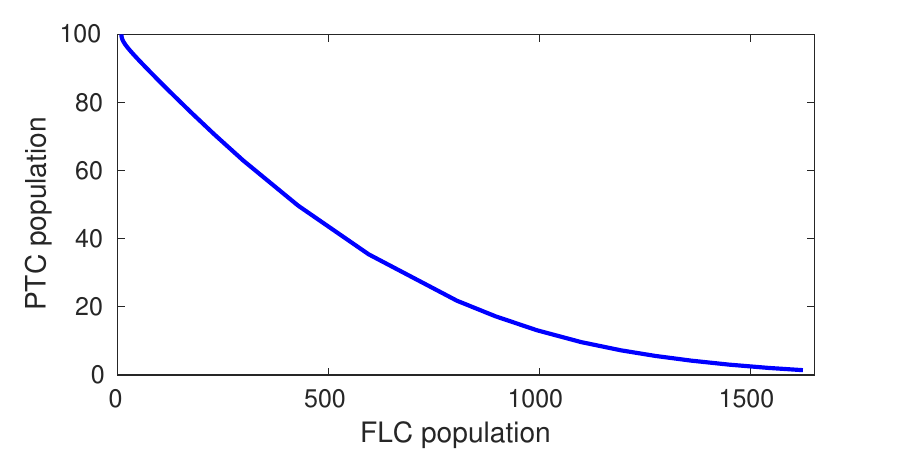}
}

\subfloat[]{
\includegraphics[width=1\textwidth, height=6cm, keepaspectratio]{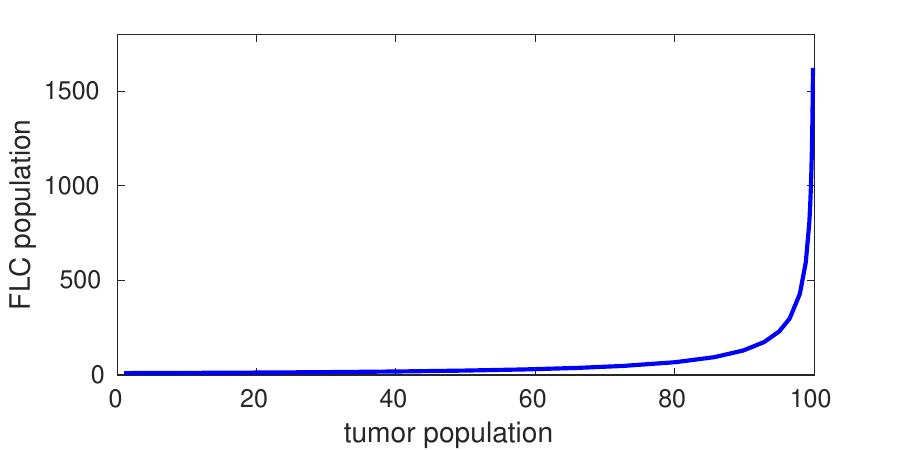}
}
\caption[Computational results for model with tumor]{Computational results for model with tumor (Equations \eqref{eq:1T}-\eqref{eq:4T}) using the parameter values in Table \ref{table:tumorparameters}, for initial conditions $P(0)=100$, $L(0)=10$, $F(0)=100$ and $T(0)=1$, over a time period of 160 days. The proximal tubule cell and tumor populations are interpreted as percentages, and free light chain population units are mg/L. This simulation represents a fast-acting tumor that is not treated, and shows the relationship between proximal tubule cells and free light chains, and the relationship between tumor growth and free light chain growth, respectively.}
\label{fig:fastPhasePlanes}
\end{figure}

Figure \ref{fig:fastPhasePlanes} demonstrates the relationships between two different sets of variables by plotting the PTC population against the FLC population in \ref{fig:fastPhasePlanes} (a) and the FLC population against the myeloma tumor (or monoclonal myeloma protein) population in \ref{fig:fastPhasePlanes} (b). Figure \ref{fig:fastPhasePlanes} (a) shows results for the model run over 160 days. As the FLC population increases, the PTC population decreases. When the FLC population reaches 500 mg/L, a common threshold for patients with kidney damage, the PTC population has decreased to between 40-50$\%$ of the original population. Figure \ref{fig:fastPhasePlanes} (b) demonstrates that as the tumor population increases, there is a delay in the response of increase to the free light chain population.

The computational results for this model simulate the expected biological relationships between the key populations involved in renal impairment in patients with MM and lays a foundation for a model incorporating the effects of treatment. Because of risks involved, patients presenting with severe renal impairment are often excluded from treatment trials. \cite{Yadav2015} An \emph{in silico} experiment via computer simulations of a mathematical model could give researchers more opportunities to investigate the impact of renal impairment on overall median survival time and patient prognosis.

\section*{Conclusions}
We have developed a mathematical model for the major cell and protein populations involved in proximal tubule cell injury due to free light chains produced by myeloma cancer cells. The system of ODEs captures the qualitative behavior of the cell populations based on a biological understanding of the process taking place inside the kidneys, that is, that the proximal tubule cells in the kidney undergo apoptosis as a result of increased production of free light chains by myeloma cells and inflammatory response involving renal fibroblasts. 

This model is a mathematical representation of the processes that take place in patients with MM that present with renal injury at diagnosis. Several simplifications were made when developing the model. These include the decision to focus on only four populations within the body and hence on parameters that reflect the effects of those populations on each other, but not other influences. In addition, this model does not take into account other factors that may influence renal function including other co-existing diseases such as diabetes mellitus and hypertension or medications.

Further work in this area includes parameterization of the theoretical model utilizing relevant markers of renal function and myeloma burden from patients suffering from this disease. It is our hope that this model will lead to further models, incorporating essential processes as needed, that could be calibrated with training sets from clinical data, and then validated by comparing model predictions against actual patient outcomes. Increasingly faithful models could be used to further investigate the relationship between renal function and serum FLC levels at presentation, or to predict the likelihood of renal function recovery following myeloma therapy for those patients presenting in renal failure. Even for patients without overt signs of renal failure it has been assumed that upon diagnosis of myeloma, some degree of renal injury has already been sustained. These modeling efforts could lead to a better understanding of this sub-clinical damage and therefore have an impact on clinical strategies which could be employed to ensure optimal renal function. Success would provide clinicians with a valuable tool with genuine prognostic and predictive capabilities.

\bibliography{library}


\begin{figure}
\renewcommand\arraystretch{0.25}
\tikzstyle{block} = [rectangle, draw, fill=blue!20, 
    text width=9.5em, text centered, rounded corners, minimum height=4em]
\tikzstyle{block2} = [rectangle, draw, fill=yellow!20, text width =8em, text centered, rounded corners, minimum height=4em]
\tikzstyle{line} = [draw, -latex']
\tikzstyle{cloud} = [draw, ellipse,fill=red!20, text width=8em, text centered, minimum height=4em]
\begin{center}
\linespread{1}
\begin{tikzpicture}[node distance = 3cm, auto]
    \node [block] (init) {Increased FLC production by myeloma cells};
    \node [block, below of=init] (endo) {Increased light chain endocytosis by PTCs via cubulin/megalin};
    \node [block2, right of=endo, node distance=4.5cm] (non) {Precipitation of non-endocytosed FLC};
    \node [block2, below of=non] (casts) {Tubular casts};
    \node [block2, below of=casts] (obstruct) {Upstream obstruction};
    \node [block, below of=endo] (act) {NF-$\kappa$B, MAPK activation in PTCs};
    \node [block, below of=act] (cyto) {IL-6, CCL2, IL-8, TGF-$\beta$ production\\ by PTCs};
    \node [cloud, left of=cyto, node distance=5.5cm] (EMT) {Proinflammatory and fibrotic\\ pathways, EMT};
    \node [block, below of=cyto] (fibro) {Renal interstitial\\ fibrosis, PTC apoptosis};
    \node [block, below of=fibro] (fail) {Chronic renal failure};
    \path [line] (init) -- (endo);
    \path [line] (init) -| (non);
    \path [line] (endo) -- (act);
    \path [line] (act) -- (cyto);
    \path [line] (cyto) -- (fibro);
    \path [line] (fibro) -- (fail);
    \path [line] (non) -- (casts);
    \path [line] (casts) -- (obstruct);
    \path [line, dashed] (cyto)--(EMT);
    \path [line, dashed] (EMT)|-(fibro);
    \path [line] (obstruct) |- (fibro);
\end{tikzpicture}
\end{center}
\caption[Flowchart of light-chain mediated renal damage in multiple myeloma] {Flowchart of light-chain mediated renal damage in multiple myeloma. Our model focuses on the progression demonstrated by the center arrows, that is, that increased FLC production by myeloma cells leads to increased light chain endocytosis by PTCs. This begins a cascade of inflammatory response mechanisms in the kidney that eventually lead to renal interstitial fibrosis, PTC apoptosis, and if untreated, chronic renal failure.}
\label{fig:Flowchart}
\end{figure}
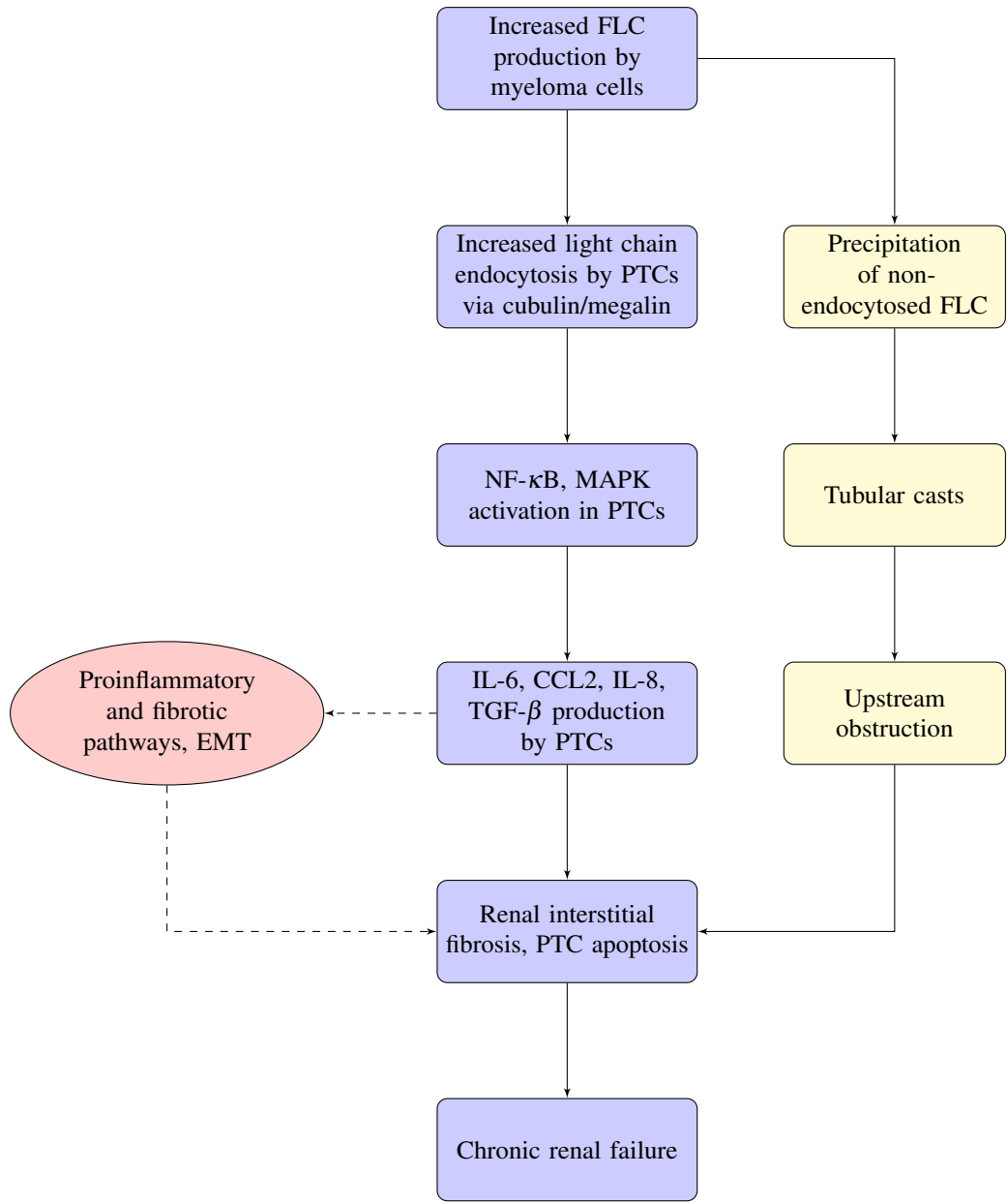 \renewcommand{\arraystretch}{0.5}

\newpage

\renewcommand{\arraystretch}{1}
\begin{table}[p]
\captionsetup{width=0.83\textwidth}
\centering
\begin{tabular}{|c|c|}
\hline
$\mathbf{(P^*, L^*, F^*)}$ & \textbf{Requirements for Stability}\\
\hline
\hline
$\left(\left(\frac{\mu_P}{\beta_P} \left( 1- \frac{L}{L_S} \right)^{-g_1}\right)^{\frac{1}{g_2-1}}, L_{\text{min}}, F_{\text{max}}\right)$ & $\gamma_F > \mu_F, \gamma_L > \mu_L, g_2 < 1$\\
$\left(\left(\frac{\mu_P}{\beta_P} \left( 1- \frac{L}{L_S} \right)^{-g_1}\right)^{\frac{1}{g_2-1}},L_{\text{min}}, 0 \right)$ & $\gamma_F < \mu_F, \gamma_L > \mu_L, g_2 < 1$\\
$\left(\left( \frac{\beta_P}{\mu_P} \right) ^{\frac{1}{1-g_2}}, 0, F_{\text{max}}\right)$ & $\gamma_F > \mu_F, \gamma_L < \mu_L, g_2 < 1$\\
$\left(\left( \frac{\beta_P}{\mu_P} \right) ^{\frac{1}{1-g_2}}, 0, 0 \right)$ & $\gamma_F < \mu_F, \gamma_L < \mu_L, g_2 < 1$\\
$(0, L_{\text{min}}, F_{\text{max}})$ & $\gamma_F > \mu_F, \gamma_L > \mu_L, \mu_P > 0$\\
$(0, L_{\text{min}}, 0)$ & $\gamma_F < \mu_F, \gamma_L > \mu_L, \mu_P > 0$ \\
$(0, 0, F_{\text{max}})$ & $\gamma_F > \mu_F, \gamma_L < \mu_L, \mu_P > 0$\\
$(0, 0, 0)$ & $\gamma_F < \mu_F, \gamma_L < \mu_L, \mu_P > 0$\\
\hline
\end{tabular}
\caption[Requirements for stability of equilibrium points]{Summary of the results of steady-state analysis on the model described by Equations \eqref{eq:1}-\eqref{eq:3}. Because we expect a healthy patient to have non-zero amounts of PTCs (P), FLCs (L), and renal fibroblasts (F), only one of these eight steady-state solutions is both mathematically and biologically significant: $(P^*, L^*, F^*) = \left(\left(\frac{\mu_P}{\beta_P} \left( 1- \frac{L}{L_S} \right)^{-g_1}\right)^{\frac{1}{g_2-1}}, L_{\text{min}}, F_{\text{max}}\right)$, so we use this steady-state solution and its requirements for stability to scale parameters for computational analysis and results.}
\label{table:genstability}
\end{table}


\begin{singlespace}
\begin{table} 
\begin{center}
\begin{tabular}{|C{5.5cm}|C{1.5cm}|C{3.3cm}|C{2.5cm}|}
\hline
 & & &  \\
Parameter name & symbol & value & units\\
 & & & \\
\hline
\hline
 & & & \\
FLC growth constant & $\gamma_L$ & .005 & 1/(\% cells  days)\\
 & & & \\
Tumor growth constant & $\gamma_T$ & .05 & 1/days\\ 
 & & & \\
Fibroblast growth constant & $\gamma_F$ & .004 & L/(mg  days)\\
 & & & \\ 
PTC proliferation constant & $\beta_P$ & .055 & 1/days\\
 & & & \\
FLC saturation constant & $L_S$ & 1000 & mg/L\\
 & & & \\
Minimum FLC level & $L_{\text{min}}$ & 10 & mg/L\\
 & & & \\
Maximum fibroblast population & $F_{\text{max}}$ & 100 & percentage \\
 & & & \\
Tumor saturation level & $T_S$ & 100 & percentage\\
 & & & \\
PTC natural apoptosis rate & $\mu_P$ & .045 & 1/days\\
 & & & \\
Natural FLC apoptosis rate & $\mu_L$ & .0005 & 1/days \\
 & & & \\
Natural fibroblast apoptosis rate& $\mu_F$ & .00046 & 1/days\\
 & & & \\
Strength of inhibited FLC on PTC growth & $g_1$ & 1 & dimensionless\\
 & & & \\
Strength of PTC on its own growth & $g_2$ & 0.958607314841775 & dimensionless\\
 & & & \\
Strength of FLC on PTC transition & $g_3$ & 0.5 & dimensionless\\
 & & & \\
Strength of PTC on PTC transition & $g_4$ & 0.38 & dimensionless\\
 & & & \\
Strength of tumor on FLC growth & $g_5$ & 0.8 & dimensionless\\
 & & & \\
Strength of FLC on its own growth & $g_6$ & 0.7 & dimensionless\\
 & & & \\
\hline
\end{tabular}
\end{center}
\caption{Parameters used to find computational results for Equations \eqref{eq:1T}-\eqref{eq:4T}. Parameters for the Gompertz tumor model ($\gamma_T$, $L_T$) are scaled to a dimensionless model, e.g. maximum tumor load is set to be 100 (representing $100 \%$ spread by tumor). Actual parameter values were taken from published work by Ayati et. al.\cite{Ayati2010} All other parameters were obtained for the model using heuristic parameter estimation to a generic response. $L_{min} = 10$ mg/L represents the expected amount of FLCs in any patient, and $L_S = 1000$ is the threshold at which renal impairment is typically observed.}
\label{table:tumorparameters}
\end{table}
\end{singlespace}


\end{document}